\documentclass{article}
\usepackage{spconf,amsmath,graphicx,hyperref}
\usepackage{booktabs}
\usepackage{multirow}

\title{THE ICASSP 2026 AUTOMATIC SONG AESTHETICS EVALUATION CHALLENGE}

\name{
Guobin Ma$^{1}$
Yuxuan Xia$^{1}$
Jixun Yao$^{1}$
Huixin Xue$^{2}$
Hexin Liu$^{3}$
Shuai Wang$^{4}$
Hao Liu$^{2}$
Lei Xie$^{1}\sthanks{Corresponding author.}$
}
\address{
$^{1}$ Audio, Speech and Language Processing Group (ASLP@NPU), School of Computer Science, \\
Northwestern Polytechnical University, Xi’an, China \\
$^{2}$ Shanghai Conservatory of Music, Shanghai, China \\
$^{3}$ College of Computing and Data Science, Nanyang Technological University, Singapore \\
$^{4}$ Nanjing University, Suzhou, China \\
}

\begin{document}
\ninept
\maketitle

\begin{abstract}

This paper summarizes the ICASSP 2026 Automatic Song Aesthetics Evaluation (ASAE) Challenge\footnote{\url{https://aslp-lab.github.io/Automatic-Song-Aesthetics-Evaluation-Challenge/}}, which focuses on predicting the subjective aesthetic scores of AI-generated songs. The challenge consists of two tracks: Track 1 targets the prediction of the overall musicality score, while Track 2 focuses on predicting five fine-grained aesthetic scores. The challenge attracted strong interest from the research community and received numerous submissions from both academia and industry. Top-performing systems significantly surpassed the official baseline, demonstrating substantial progress in aligning objective metrics with human aesthetic preferences. The outcomes establish a standardized benchmark and advance human-aligned evaluation methodologies for modern music generation systems.
\end{abstract}


\vspace{-5pt}
\section{Introduction}
\label{sec:intro}

With the rapid development of generative AI, music generation has progressed from producing merely coherent audio waveforms to creating artistic content that evokes emotional resonance in listeners. 
Commercial music generators (e.g., Suno~\cite{Suno}, Udio~\cite{Udio}, and Mureka~\cite{Mureka}) have transformed the composition process by generating high-fidelity songs and narrowing the gap between human and machine creativity. However, evaluating the subjective aesthetic of generated songs, capturing aspects such as emotional expressiveness, musicality, and listener enjoyment, remains a significant challenge. Objective evaluation metrics such as Fréchet Audio Distance~\cite{DBLP:conf/interspeech/KilgourZRS19} capture distributional distance but weakly correlate with aesthetic perception~\cite{DBLP:conf/ismir/Vinay022,DBLP:conf/eusipco/TailleurLLCHIO24}. Reference-free models like Meta Audiobox Aesthetics~\cite{DBLP:journals/corr/abs-2502-05139} generalize across audio domains but prioritize acoustic fidelity over high-level musical aesthetics.

To address this gap, we organized the ICASSP 2026 ASAE Challenge, aiming to establish a standardized benchmark for reference-free aesthetics evaluation of generated songs, focusing on human listener-centered criteria. The challenge comprises two tracks: Overall Musicality Prediction~(track 1), which focuses on a holistic assessment, and Fine-grained Dimension Prediction~(track 2), which extends the evaluation to five specific aesthetic dimensions.

\vspace{-5pt}
\section{Dataset}
\label{sec:dataset}

To support this challenge, we adopt the SongEval benchmark~\cite{yao2025songeval}, using its dataset for training and its UTMOS-based~\cite{DBLP:conf/interspeech/SaekiXNKTS22} model as the official baseline\footnote{\url{https://github.com/ASLP-lab/SongEval}}. We further construct dedicated test sets to rigorously evaluate model performance, where all songs were annotated by experts with formal musical backgrounds.

Track 1 includes two evaluation subsets: a \textit{\textbf{Regular}} set containing songs generated by 5 seen systems (Suno~\cite{Suno}, Udio~\cite{Udio}, Mureka~\cite{Mureka}, DiffRhythm~\cite{ning2025diffrhythm}, YUE~\cite{DBLP:journals/corr/yue}) to assess in-distribution performance, and a \textit{\textbf{Hard}} set comprising the 5 seen plus 6 unseen systems (Seed-Music~\cite{Seed-Music}, ElevenLabs~\cite{ElevenLabs}, ACE-step~\cite{DBLP:journals/corr/acestep}, JAM~\cite{DBLP:journals/corr/abs-jam}, SongGeneration~\cite{DBLP:journals/corr/levo}, Songbloom~\cite{DBLP:journals/corr/abs-songbloom}) to evaluate robust generalization.

Track 2 uses a purpose-built, curated test set targeting five aesthetic dimensions: \textit{Overall Coherence}, \textit{Memorability}, \textit{Vocal Naturalness}, \textit{Structure Clarity}, and \textit{Overall Musicality}. This dataset ensures broad coverage of aesthetic qualities and includes samples with isolated degradations (e.g., high \textit{Memorability} but low \textit{Coherence}), enabling fine-grained evaluation of dimension-specific sensitivity.

\begin{figure*}[!ht]
  \centering
  \includegraphics[width=\textwidth]{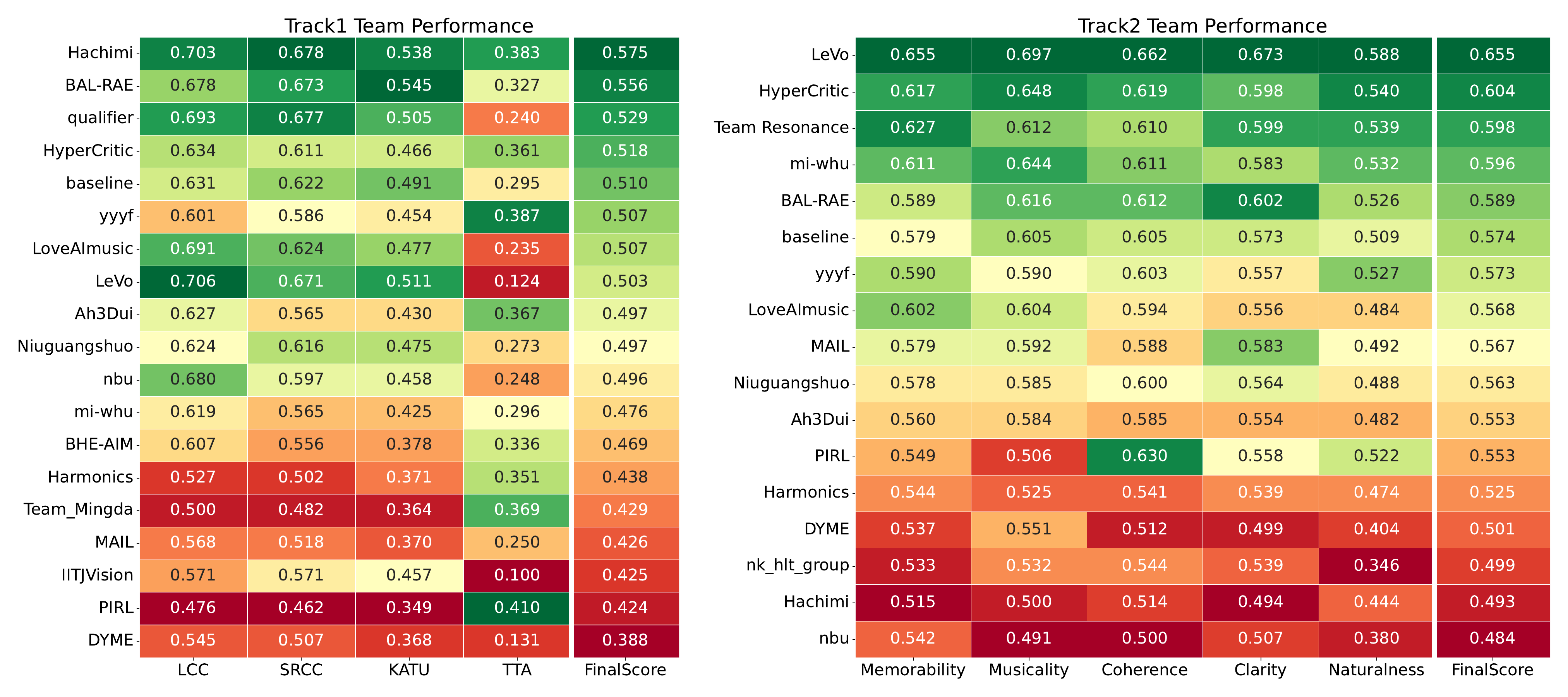}
  \vspace{-13pt}
  \caption{Overview of the challenge results. The visualizations depict the performance distribution of participating teams in Track 1 (Overall Musicality) and Track 2 (Fine-grained Dimensions) compared to the baseline system.}
  \label{fig:main_results}
  \vspace{-8pt}
\end{figure*}

\vspace{-5pt}
\section{Challenge Metrics}
\label{sec:tracks}

We propose a hierarchical evaluation metric that assesses performance at both the \textit{Utterance-level} (individual songs) and \textit{System-level} (averaged scores per model).

\vspace{-5pt}
\subsection{Track 1: Overall Musicality Prediction}
Track 1 focuses on predicting a single scalar score representing the global aesthetic quality. To test model generalization, we employed two distinct test sets described in Sec.~\ref{sec:dataset}: \textbf{Set 1 (\textit{Regular})} and \textbf{Set 2 (\textit{Hard})}.
For each test set, we calculate a composite score, $S_{set}$, by aggregating four metrics: Pearson Linear Correlation Coefficient (LCC)~\cite{sedgwick2012pearson}, Spearman Rank Correlation Coefficient (SRCC)~\cite{sedgwick2014spearman}, Kendall’s Tau (KATU)~\cite{mcleod2005kendall}, and Top-Tier Accuracy (TTA\footnote{TTA measures the accuracy in correctly classifying the top-tier songs.}). Note that TTA is calculated only at the utterance level. The composite score is defined as:
\begin{equation}
\label{eq:set_score}
S_{set} = \frac{1}{4} \left( \text{TTA}_{u} + \sum_{m \in \mathcal{M}} \frac{m_{u} + m_{s}}{2} \right)
\end{equation}
where $\mathcal{M} = \{\text{LCC, SRCC, KATU}\}$, and subscripts $u$ and $s$ denote the utterance and system levels, respectively.

To encourage robustness, the final ranking for Track 1 is determined by a weighted sum prioritizing the Hard Set:
\begin{equation}
\text{Score}_{\text{Track1}} = 0.2 \times S_{\text{set1}} + 0.8 \times S_{\text{set2}}
\end{equation}

\vspace{-10pt}
\subsection{Track 2: Fine-grained Dimension Prediction}

Track 2 requires participants to predict scores for all five aesthetic dimensions simultaneously. Evaluation is performed on the dedicated Track 2 test set described in Sec.~\ref{sec:dataset}.

For each dimension $d \in \{1, \dots, 5\}$, we calculate a dimensional score, $S_{d}$, utilizing the same composite metric formulation as Eq.~(\ref{eq:set_score}).
The final ranking metric for Track 2 is the macro-average across all aesthetic dimensions:
\begin{equation}
\label{eq:track2_score}
\text{Score}_{\text{Track2}} = \frac{1}{5} \sum_{d=1}^{5} S_{d}
\end{equation}

This protocol favors models that demonstrate consistent performance across all aesthetic aspects, rather than those biased towards specific dimensions.

\vspace{-5pt}
\section{Results and Conclusion}
\label{sec:results}

The challenge received substantial attention from the community, with over 70 teams registering. Ultimately, we received 18 and 16 valid submissions for Track 1 and Track 2, respectively. Fig.~\ref{fig:main_results} summarizes the performance of participating teams in both tracks.

\textbf{For Track 1}, Top-performing systems consistently surpassed the official baseline, and most converged on a common strategy: leveraging multiple pretrained audio representations and explicitly modeling long-range song structure. 
The winning team, \textit{\textbf{Hachimi}}, fused MuQ \cite{DBLP:journals/corr/abs-muq-01108} and WavLM \cite{DBLP:journals/jstsp/ChenWCWLCLKYXWZ22} embeddings using two-stage bidirectional cross-attention modules, enabling mutual information exchange between music-structural cues and acoustic details, which is particularly beneficial for holistic musicality prediction. 
\textit{\textbf{BAL-RAE}} enhanced robustness through multi-scale feature extraction and a hybrid ranking-aware optimization strategy, aligning the training objective with the competition's correlation and TTA metrics. 
\textit{\textbf{Qualifier}} combined large-scale music pretraining with a metric-based meta-learning framework, which effectively mitigated overfitting under limited labels and improved generalization to the Hard split.
\textit{\textbf{HyperCritic}} also performed strongly by integrating heterogeneous SSL backbones with a structure-aware branch (e.g., SongFormer \cite{hao2025songformer}), further validating the importance of structural modeling for global aesthetics.

\vspace{4pt}
\textbf{For Track 2}, strong entries emphasized dimension-specific modeling and balanced performance across attributes. 
The winning team, \textit{\textbf{LeVo}}, adopted a mixture-of-experts fusion strategy over diverse pretrained backbones. This approach enabled the adaptive weighting of complementary cues—ranging from acoustic to generative features—achieving the most consistent scores across the five dimensions. 
The next best teams, including \textit{\textbf{HyperCritic}} and \textit{\textbf{Team Resonance}}, further highlight the benefit of diversity: the former integrated multiple feature streams via a unified transformer \cite{DBLP:conf/nips/VaswaniSPUJGKP17}, while the latter employed a robust regression ensemble to reduce variance. 
The \textit{\textbf{mi-whu}} team achieved competitive results with a dual-encoder design that shares common musical factors while retaining dimension-specific pooling. Coupled with data augmentation and lightweight fine-tuning, this strategy strengthens fine-grained sensitivity without sacrificing stability.

\vspace{4pt}
Collectively, these outcomes indicate clear progress in aligning objective metrics with human aesthetic judgment. Top systems highlight the effectiveness of multi-representation fusion, structure-aware modeling, and correlation-aligned objectives. 
However, a persistent bottleneck remains: despite strong correlation trends, TTA scores across teams were generally modest. This discrepancy suggests that while models successfully capture global aesthetic rankings, they struggle with fine-grained discrimination at the high-quality end, frequently under-scoring top-tier tracks. 
We hope this challenge will accelerate advances in music generation, while the accompanying evaluation models can support system optimization and serve as reward functions in reinforcement learning frameworks.

\vfill\pagebreak

\bibliographystyle{IEEEbib}
\bibliography{strings,refs}

@inproceedings{DBLP:conf/ismir/Vinay022,
  author       = {Ashvala Vinay and
                  Alexander Lerch},
  title        = {Evaluating Generative Audio Systems and Their Metrics},
  booktitle    = {{ISMIR}},
  pages        = {858--865},
  year         = {2022}
}

@inproceedings{DBLP:conf/eusipco/TailleurLLCHIO24,
  author       = {Modan Tailleur and
                  Junwon Lee and
                  Mathieu Lagrange and
                  Keunwoo Choi and
                  Laurie M. Heller and
                  Keisuke Imoto and
                  Yuki Okamoto},
  title        = {Correlation of Fr{\'{e}}chet Audio Distance With Human Perception
                  of Environmental Audio Is Embedding Dependent},
  booktitle    = {{EUSIPCO}},
  pages        = {56--60},
  publisher    = {{IEEE}},
  year         = {2024}
}

@inproceedings{DBLP:conf/interspeech/KilgourZRS19,
  author       = {Kevin Kilgour and
                  Mauricio Zuluaga and
                  Dominik Roblek and
                  Matthew Sharifi},
  title        = {Fr{\'{e}}chet Audio Distance: {A} Reference-Free Metric for Evaluating
                  Music Enhancement Algorithms},
  booktitle    = {{INTERSPEECH}},
  pages        = {2350--2354},
  publisher    = {{ISCA}},
  year         = {2019}
}

@inproceedings{DBLP:conf/interspeech/SaekiXNKTS22,
  author       = {Takaaki Saeki and
                  Detai Xin and
                  Wataru Nakata and
                  Tomoki Koriyama and
                  Shinnosuke Takamichi and
                  Hiroshi Saruwatari},
  title        = {{UTMOS:} UTokyo-SaruLab System for VoiceMOS Challenge 2022},
  booktitle    = {{INTERSPEECH}},
  pages        = {4521--4525},
  publisher    = {{ISCA}},
  year         = {2022}
}

@inproceedings{DBLP:conf/nips/VaswaniSPUJGKP17,
  author       = {Ashish Vaswani and
                  Noam Shazeer and
                  Niki Parmar and
                  Jakob Uszkoreit and
                  Llion Jones and
                  Aidan N. Gomez and
                  Lukasz Kaiser and
                  Illia Polosukhin},
  title        = {Attention is All you Need},
  booktitle    = {{NIPS}},
  pages        = {5998--6008},
  year         = {2017}
}

@article{DBLP:journals/corr/abs-2502-05139,
  author       = {Andros Tjandra and
                  Yi{-}Chiao Wu and
                  Baishan Guo and
                  John Hoffman and
                  Brian Ellis and
                  Apoorv Vyas and
                  Bowen Shi and
                  Sanyuan Chen and
                  Matt Le and
                  Nick Zacharov and
                  Carleigh Wood and
                  Ann Lee and
                  Wei{-}Ning Hsu},
  title        = {Meta Audiobox Aesthetics: Unified Automatic Quality Assessment for
                  Speech, Music, and Sound},
  journal      = {CoRR},
  volume       = {abs/2502.05139},
  year         = {2025}
}

@article{DBLP:journals/corr/yue,
  author       = {Ruibin Yuan and
                  Hanfeng Lin and
                  Shuyue Guo and
                  Ge Zhang and
                  Jiahao Pan and
                  Yongyi Zang and
                  Haohe Liu and
                  Yiming Liang and
                  Wenye Ma and
                  Xingjian Du and
                  Xinrun Du and
                  Zhen Ye and
                  Tianyu Zheng and
                  Yinghao Ma and
                  Minghao Liu and
                  Zeyue Tian and
                  Ziya Zhou and
                  Liumeng Xue and
                  Xingwei Qu and
                  Yizhi Li and
                  Shangda Wu and
                  Tianhao Shen and
                  Ziyang Ma and
                  Jun Zhan and
                  Chunhui Wang and
                  Yatian Wang and
                  Xiaowei Chi and
                  Xinyue Zhang and
                  Zhenzhu Yang and
                  Xiangzhou Wang and
                  Shansong Liu and
                  Lingrui Mei and
                  Peng Li and
                  Junjie Wang and
                  Jianwei Yu and
                  Guojian Pang and
                  Xu Li and
                  Zihao Wang and
                  Xiaohuan Zhou and
                  Lijun Yu and
                  Emmanouil Benetos and
                  Yong Chen and
                  Chenghua Lin and
                  Xie Chen and
                  Gus Xia and
                  Zhaoxiang Zhang and
                  Chao Zhang and
                  Wenhu Chen and
                  Xinyu Zhou and
                  Xipeng Qiu and
                  Roger B. Dannenberg and
                  Zheng{-}Jia Liu and
                  Jian Yang and
                  Wenhao Huang and
                  Wei Xue and
                  Xu Tan and
                  Yike Guo},
  title        = {YuE: Scaling Open Foundation Models for Long-Form Music Generation},
  journal      = {CoRR},
  volume       = {abs/2503.08638},
  year         = {2025}
}

@article{yao2025songeval,
  title   = {SongEval: A Benchmark Dataset for Song Aesthetics Evaluation},
  author  = {Jixun Yao and
             Guobin Ma and
             Huixin Xue and
             Huakang Chen and
             Chunbo Hao and
             Yuepeng Jiang and
             Haohe Liu and
             Ruibin Yuan and
             Jin Xu and
             Wei Xue and
             others},
  journal = {abs/2505.10793},
  year    = {2025}
}

@article{ning2025diffrhythm,
  title   = {DiffRhythm: Blazingly fast and embarrassingly simple end-to-end full-length song generation with latent diffusion},
  author  = {Ziqian Ning and
             Huakang Chen and
             Yuepeng Jiang and
             Chunbo Hao and
             Guobin Ma and
             Shuai Wang and
             Jixun Yao and
             Lei Xie},
  journal = {abs/2503.01183},
  year    = {2025}
}

@article{DBLP:journals/corr/acestep,
  author       = {Junmin Gong and
                  Sean Zhao and
                  Sen Wang and
                  Shengyuan Xu and
                  Joe Guo},
  title        = {ACE-Step: {A} Step Towards Music Generation Foundation Model},
  journal      = {CoRR},
  volume       = {abs/2506.00045},
  year         = {2025}
}

@article{DBLP:journals/corr/levo,
  author       = {Shun Lei and
                  Yaoxun Xu and
                  Zhiwei Lin and
                  Huaicheng Zhang and
                  Wei Tan and
                  Hangting Chen and
                  Jianwei Yu and
                  Yixuan Zhang and
                  Chenyu Yang and
                  Haina Zhu and
                  Shuai Wang and
                  Zhiyong Wu and
                  Dong Yu},
  title        = {LeVo: High-Quality Song Generation with Multi-Preference Alignment},
  journal      = {CoRR},
  volume       = {abs/2506.07520},
  year         = {2025}
}

@article{DBLP:journals/corr/abs-jam,
  author       = {Renhang Liu and
                  Chia{-}Yu Hung and
                  Navonil Majumder and
                  Taylor Gautreaux and
                  Amir Ali Bagherzadeh and
                  Chuan Li and
                  Dorien Herremans and
                  Soujanya Poria},
  title        = {{JAM:} {A} Tiny Flow-based Song Generator with Fine-grained Controllability
                  and Aesthetic Alignment},
  journal      = {CoRR},
  volume       = {abs/2507.20880},
  year         = {2025}
}

@article{DBLP:journals/corr/abs-songbloom,
  author       = {Chenyu Yang and
                  Shuai Wang and
                  Hangting Chen and
                  Wei Tan and
                  Jianwei Yu and
                  Haizhou Li},
  title        = {SongBloom: Coherent Song Generation via Interleaved Autoregressive
                  Sketching and Diffusion Refinement},
  journal      = {CoRR},
  volume       = {abs/2506.07634},
  year         = {2025}
}

@article{sedgwick2012pearson,
  title     = {Pearson’s correlation coefficient},
  author    = {Philip Sedgwick},
  journal   = {Bmj},
  volume    = {345},
  year      = {2012},
  publisher = {British Medical Journal Publishing Group}
}

@article{sedgwick2014spearman,
  title     = {Spearman’s rank correlation coefficient},
  author    = {Philip Sedgwick},
  journal   = {Bmj},
  volume    = {349},
  year      = {2014},
  publisher = {British Medical Journal Publishing Group}
}

@article{mcleod2005kendall,
  title   = {Kendall rank correlation and Mann-Kendall trend test},
  author  = {A Ian McLeod},
  journal = {R package Kendall},
  volume  = {602},
  pages   = {1--10},
  year    = {2005}
}

@article{DBLP:journals/corr/abs-muq-01108,
  author       = {Haina Zhu and
                  Yizhi Zhou and
                  Hangting Chen and
                  Jianwei Yu and
                  Ziyang Ma and
                  Rongzhi Gu and
                  Yi Luo and
                  Wei Tan and
                  Xie Chen},
  title        = {MuQ: Self-Supervised Music Representation Learning with Mel Residual
                  Vector Quantization},
  journal      = {CoRR},
  volume       = {abs/2501.01108},
  year         = {2025}
}

@article{DBLP:journals/jstsp/ChenWCWLCLKYXWZ22,
  author       = {Sanyuan Chen and
                  Chengyi Wang and
                  Zhengyang Chen and
                  Yu Wu and
                  Shujie Liu and
                  Zhuo Chen and
                  Jinyu Li and
                  Naoyuki Kanda and
                  Takuya Yoshioka and
                  Xiong Xiao and
                  Jian Wu and
                  Long Zhou and
                  Shuo Ren and
                  Yanmin Qian and
                  Yao Qian and
                  Jian Wu and
                  Michael Zeng and
                  Xiangzhan Yu and
                  Furu Wei},
  title        = {WavLM: Large-Scale Self-Supervised Pre-Training for Full Stack Speech
                  Processing},
  journal      = {{IEEE} J. Sel. Top. Signal Process.},
  volume       = {16},
  number       = {6},
  pages        = {1505--1518},
  year         = {2022}
}

@article{hao2025songformer,
  title   = {Songformer: Scaling music structure analysis with heterogeneous supervision},
  author  = {Chunbo Hao and
             Ruibin Yuan and
             Jixun Yao and
             Qixin Deng and
             Xinyi Bai and
             Wei Xue and
             Lei Xie},
  journal = {arXiv preprint arXiv:2510.02797},
  year    = {2025}
}

@article{Suno,
  title={Suno official website},
  author={Suno},
  journal = {Suno},
  year={2025},
  note={\url{https://suno.com}}
}

@article{Udio,
  title={Udio official website},
  author={Udio},
  journal = {Udio},
  year={2025},
  note={\url{https://www.udio.com/home}}
}

@article{Mureka,
  title={Mureka official website},
  author={Mureka},
  journal = {Mureka},
  year={2025},
  note={\url{https://www.mureka.ai}}
}

@article{Seed-Music,
  title={Seed-Music official website},
  author={Seed-Music},
  journal = {Seed-Music},
  year={2025},
  note={\url{https://www.doubao.com/chat}}
}

@article{ElevenLabs,
  title={ElevenLabs official website},
  author={ElevenLabs},
  journal = {ElevenLabs},
  year={2025},
  note={\url{https://elevenlabs.io/app/music}}
}

\end{document}